\documentclass[12pt]{article}
\usepackage[english]{babel}
\usepackage{latexsym}
\usepackage{amsfonts}
\begin{document}
\begin{center}
{\Large \bf The strong coupling effect and auxiliary fields\\ in
the DGP-model}\\

\vspace{4mm}

Mikhail N.~Smolyakov\\

\vspace{4mm} Skobeltsyn Institute of Nuclear Physics, Moscow State
University,\\ Vorob'evy Gory, 119992 Moscow, Russia\\
\end{center}

\begin{abstract}
The DGP-model with additional terms in the action is considered.
These terms have a special form and include auxiliary scalar
fields without kinetic terms, which are non-minimally coupled to
gravity. The use of these fields allows one to exclude the mode,
which corresponds to the strong coupling effect, from the theory.
Effective four-dimensional theory on the brane appears to be the
same, as in the original DGP-model.
\end{abstract}

\section{Introduction}
The model proposed in \cite{DGP} possesses a remarkable feature --
a modification of gravity at ultra-large scales. This effect is
very interesting from the cosmological point of view, namely for
possible explanation of the observed large-scale acceleration of
our Universe \cite{cosmo}. Unfortunately the strong coupling
effect due to the existence of a strongly interacting mode, which
is a four-dimensional scalar, was found in the model
\cite{Luty,Rub}. This problem was discussed in a lot of papers and
there were proposed some mechanisms of its solution. For example,
it was argued that the strong coupling effect is an artifact of
linear approximation and disappears after taking into account the
non-linear effects (resummation of perturbative series)
\cite{Infrared}. Another approach is to utilize a regularization
in the model \cite{KPR,PR}. In this paper a new mechanism for
overcoming the strong coupling, based on a slight modification of
the original action, is proposed.

The paper is organized as follows. First, we solve exactly
equations for linearized gravity in the DGP-model and explicitly
distinguish the strong coupled mode, which is nothing else but the
$h_{44}$-component of metric fluctuations. This result is
well-known and was mentioned earlier, for example, in
\cite{DubLib}. Then we show that with the help of additional terms
in the action, which contain "auxiliary"\ scalar fields
non-minimally interacting with gravity, one can exclude this
strongly interacting mode from the theory. This is shown by
solving the corresponding equations of motion and identifying the
physical degrees of freedom of the model. At the same time
effective equation for the $\mu\nu$-component of metric
fluctuations remains unchanged and one can use all
phenomenological predictions obtained for gravity on the brane.
Finally we briefly discuss the obtained results and possible
consequences.

\section{Gravity in the original DGP-model}
Action of the model proposed in \cite{DGP} has the following form:
\begin{equation}\label{DGPaction}
S_{DGP}=M_{*}^{3}\int R\sqrt{-g}\,d^{4}xdy + \Omega
M_{*}^{3}\int_{y=0} \tilde R\sqrt{-\tilde g}\,d^{4}x,
\end{equation}
where $M_{*}$ is the five-dimensional Planck mass,
$\Omega>>1/M_{*}$ and $\tilde g_{\mu\nu}$ is the induced metric on
the brane, which is located at the point $y=0$ of the extra
dimension. In addition, let us suppose that the model possesses
$y\leftrightarrow -y$ symmetry, which fixes the brane position. We
also note that the signature of the metric $g_{MN}$ is chosen to
be $(-,+,+,+,+)$.

Let us denote $\hat \kappa = \frac{1}{\sqrt{M_{*}^{3}}}$ and
parameterize the metric $g_{MN}$ as
\begin{equation}\label{metricpar}
  g_{MN} = \gamma_{MN} + \hat \kappa h_{MN},
\end{equation}
$h_{MN}$ being the metric fluctuations and $M,N=0,..,4$.
Substituting this parameterization into (\ref{DGPaction}) and
retaining the terms of the zero order in $\hat \kappa$, we can get
the second variation action of the model. This action is invariant
under the gauge transformations
\begin{eqnarray}\label{gaugetr1}
h'_{\mu\nu}\left(x,y\right)&=&h_{\mu\nu}\left(x,y\right)-\left(
\partial_{\mu}\xi_{\nu} +\partial_{\nu}\xi_{\mu}\right),\\ \label{gaugetr2}
h'_{\mu4}\left(x,y\right)&=&h_{\mu4}\left(x,y\right)-\left(
\partial_{\mu}\xi_{4} +\partial_{4}\xi_{\mu}\right),\\ \label{gaugetr3}
h'_{44}\left(x,y\right)&=&h_{44}\left(x,y\right)-2\partial_{4}\xi_{4},
\end{eqnarray}
and the functions $\xi^M(x,y)$ satisfy the symmetry conditions
\begin{eqnarray}\label{orbifoldsym1}
\xi^{\mu}\left(x,-y\right)&=&\xi^{\mu}\left(x,y\right),\\
\nonumber \xi^{4}\left(x,-y\right)&=&-\xi^{4}\left(x,y\right).
\nonumber
\end{eqnarray}
Here $\mu,\nu=0,..,3$. The matter is assumed to be placed on the
brane only, and its interaction with gravity has the standard form
\begin{equation}\label{interaction}
 \frac{\hat \kappa}{2} \int_{y=0} h^{\mu\nu}(x,0) t_{\mu\nu}(x) d^{4}x,
\end{equation}
$t_{\mu\nu}(x)$ denoting the energy-momentum tensor of the matter.

It is necessary to emphasize that in general all fluctuations of
metric must satisfy the physical boundary conditions at
$y\to\pm\infty$, $ x^i\to\pm\infty \quad (i=1,2,3)$, i.e. vanish
at spatial infinity. This is a reasonable assumption - for
example, the $h_{00}$-component is associated with  Newton's
potential, which must vanish at infinity (for the matter, which is
localized in some finite domain). Obviously, the gauge functions
$\xi^M(x,y)$ must be finite everywhere -- it follows simply from
the definition of the gauge transformations. It means that
$\xi_M(x,y)$ must be finite too (the problem of physical boundary
conditions in the case of RS2 model was discussed in detail in
\cite{SV2}).

The equations of motion for the metric fluctuations, corresponding
to action (\ref{DGPaction}), have the following form (they can be
easily derived from the corresponding equations in \cite{SV2,SM}):
1) $\mu\nu$-component
\begin{eqnarray}\label{mu-nu}
 & &\frac{1}{2} \left(\Box h_{\mu\nu}-
\partial_\mu \partial^\rho
h_{\rho\nu}-\partial_\nu \partial^\rho h_{\rho\mu} +
\partial_4\partial_4 h_{\mu\nu}+\partial_\mu
\partial_\nu h+ \partial_\mu \partial_\nu
h_{44}\right) -\\ \nonumber
&-&\frac{1}{2}\,\partial_4(\partial_{\mu}h_{\nu
4}+\partial_{\nu}h_{\mu 4})+ \frac{1}{2}
\gamma_{\mu\nu}\left(\partial^\rho
\partial^\sigma h_{\rho\sigma}-\Box
h - \partial_4\partial_4  h - \Box h_{44} +
2\partial^{\rho}\partial_4 h_{\rho 4}\right) +\\ \nonumber &+&
\frac{\Omega}{2}\delta (y)\left[\left(\Box h_{\mu\nu}-
\partial_\mu \partial^\rho
h_{\rho\nu}-\partial_\nu \partial^\rho h_{\rho\mu}+\partial_\mu
\partial_\nu h\right)+\gamma_{\mu\nu}\left(\partial^\rho
\partial^\sigma h_{\rho\sigma}-\Box
h\right)\right]=
\\ \nonumber &=& -\frac{\hat \kappa}{2} t_{\mu\nu}(x)\delta(y),
\end{eqnarray}

 2) $\mu 4$-component
\begin{equation}\label{mu-4}
\partial_4 ( \partial_\mu h - \partial^\nu  h_{\mu\nu})-
\partial^{\nu}\left(\partial_{\mu}h_{\nu 4}-\partial_{\nu}h_{\mu 4}\right)= 0,
\end{equation}

 3) $4 4$-component
\begin{equation}\label{4-4}
\partial^\mu \partial^\nu  h_{\mu\nu} - \Box  h =0,
\end{equation}
where $h=\eta^{\mu\nu}h_{\mu\nu}$ and
$\Box=\eta^{\mu\nu}\partial_{\mu}\partial_{\nu}$.

In what follows, we will also use an auxiliary equation, which is
obtained by substituting the equation for $44$-component into the
contracted equation for $\mu\nu$-component. This equation contains
$h$, $h_{\mu 4}$ and $h_{44}$ only and has the form:
\begin{equation}\label{contracted-44}
\partial_4\partial_4 h -2\partial_4\partial^{\mu}h_{\mu 4}+
\Box h_{44} = \frac{\hat \kappa}{3}\, t_{\mu}^{\mu}(x)\delta(y),
\end{equation}
where $\Box=\eta^{\mu\nu}\partial_{\mu}\partial_{\nu}$. By
integrating this equation over $y$ in the limits
$(-\infty,\infty)$ and using the physical boundary conditions for
the fields $h_{\mu\nu}$, $h_{\mu 4}$ and $h_{44}$, we  find that
the function $\rho(x)$, defined by
\begin{equation}
\rho(x)=\int_{-\infty}^\infty h_{44}(x,y)\, dy,
\end{equation}
is not equal to zero and satisfies the equation
\begin{equation}\label{h44t}
\Box\rho(x) = \frac{\hat \kappa}{3}\,
\eta^{\mu\nu}t_{\mu\nu}(x)\equiv\frac{\hat \kappa}{3}\, t(x).
\end{equation}

This means that the field $h_{44}$ cannot be gauged out, because
otherwise the equations  of motion for linearized gravity become
inconsistent. One can easily check that in the absence of the
fields $h_{\mu 4}$ and $h_{44}$ from equation
(\ref{contracted-44}) follows that
\begin{equation}
h\sim t|y|.
\end{equation}
Thus $h$ diverges at $y=\pm\infty$ and does not satisfy the
physical boundary conditions.

We will use the following form of $\xi_4$ to impose an appropriate
gauge on the field $h_{44}$ \cite{SV2}:
\begin{equation}\label{xi4}
\xi_4(x,y)=\frac{1}{4}\int_{-y}^y h_{44} (x,y') dy' -\frac{1}{4C}
\int_{-y}^y F(y') dy' \int_{-\infty}^{\infty} h_{44} (x,y') dy',
\end{equation}
where $F|_{y\to \pm\infty}=0$ and
\begin{equation}
C=\int_{-\infty}^{\infty} F(y) dy.
\end{equation}
Note that $\xi_4$ satisfies the symmetry and the boundary
conditions. With the help of (\ref{xi4}) we can pass to the gauge,
in which
\begin{equation}\label{gaugephi}
h_{44}(x,y)=F(y)\phi(x),
\end{equation}
where
\begin{equation}
\phi(x)=\frac{1}{C}\int_{-\infty}^{\infty} h_{44} (x,y) dy
\end{equation}
and depends on $x$ only (it is evident that $\rho(x)=C\phi(x)$).
It turns out to be convenient to choose $F(y)=e^{-k|y|}$.
Obviously, the field $h_{44}$ satisfies the symmetry and the
physical boundary conditions in this gauge. Moreover, we have no
residual gauge transformations with $\xi_4$. We also note that
since $\xi_4(x,0)=0$, the brane remains unshifted in this gauge.

Now let us discuss the  gauge condition for the field $h_{\mu4}$.
Let us take the gauge function $\xi_{\mu}(x,y)$ in the following
form:
\begin{equation}\label{gaugetrMuNu}
\xi_{\mu}(x,y)=\int_{-\infty}^{y} h_{\mu 4}\left(x,y'\right)\,dy'.
\end{equation}
One can easily see that due to the symmetry $h_{\mu
4}(x,-y)=-h_{\mu 4}(x,y)$, $\xi_{\mu}(x,y)$ satisfies the symmetry
condition $\xi_{\mu}(x,-y)=\xi_{\mu}(x,y)$. Moreover, it is easy
to see that $\xi_{\mu}(x,y)|_{y\to\pm\infty}\to 0$, at least in
the sense of the principal value of the integral in equation
(\ref{gaugetrMuNu}) (again due to the symmetry of $h_{\mu 4}$),
i.e. it is finite. Finally, it is not difficult to check that the
gauge transformation with $\xi_{\mu}$ given by (\ref{gaugetrMuNu})
gauges the field $h_{\mu 4}$ out. It seems that this formal
argumentation can be used in favor of the possibility to make the
$h_{\mu 4}$-field vanish  everywhere. Anyway, as we will see
later, equations of motion can be solved in the gauge $h_{\mu 4}=
0$

After this gauge fixing we are still left with residual gauge
transformations of the form
\begin{equation}\label{remgaugetr}
\partial_{4}\xi_{\mu}=0.
\end{equation}
Using the physical boundary conditions for the field $h_{\mu\nu}$
at $y\to\infty$, from equation (\ref{gaugetr1}) it follows that
\begin{equation}
\xi_{\mu}=0.
\end{equation}
Thus, we have no residual gauge transformations with $\xi_{\mu}\ne
0$.

The substitution, which allows us to solve equations of motion in
the gauge $h_{\mu4}(x,y)= 0$, $h_{44}(x,y)=e^{-k|y|}\phi(x)$, has
the form
\begin{equation}\label{substitution}
h_{\mu\nu}=b_{\mu\nu}-\frac{1}{k^{2}}e^{-k|y|}\partial_{\mu}\partial_{\nu}\phi.
\end{equation}
Substituting (\ref{substitution}) into (\ref{mu-4}), (\ref{4-4}),
(\ref{contracted-44}) we get
\begin{equation}\label{mu-4-1}
\partial_4 ( \partial_\mu b - \partial^\nu  b_{\mu\nu})
= 0,
\end{equation}
\begin{equation}\label{4-4-1}
\partial^\mu \partial^\nu  b_{\mu\nu} - \Box b=0,
\end{equation}
\begin{equation}\label{contracted-44-1}
\partial_4\partial_4 b +\frac{2}{k}
\,\Box\,\phi \delta(y)  = \frac{\hat \kappa}{3}\, t\delta(y).
\end{equation}
Integrating (\ref{contracted-44-1}) over $y$ in the limits
$(-\infty,\infty)$ and using the physical boundary conditions for
the field $b_{\mu\nu}$, we get
\begin{equation}\label{phi}
\Box\,\phi = \frac{\hat \kappa k}{6} t.
\end{equation}
This mean that
\begin{equation}
\partial_4 b =B(x),
\end{equation}
where $B(x)$ is some function of $x$ only. Using the symmetry
conditions, we obtain $B(x)\equiv 0$.

From the physical boundary conditions it follows that
\begin{equation}\label{T}
b =0,
\end{equation}
\begin{equation}\label{gaugecond}
 \partial^\mu  b_{\mu\nu} = 0.
\end{equation}
where $b=\eta^{\mu\nu}b_{\mu\nu}$. It is not difficult to find
corresponding equation for the field $b_{\mu\nu}$.

It is evident that such coupling constant for the field $\phi$
makes $\hat\kappa h_{44}$ to be much larger than unity if
$\Omega>>1/M_{*}$ (for example, in the case $M_{*}\sim 10^{-3}
eV$) at least at $y=0$. Linearized approximation breaks down, and
this is the origin of the so-called strong coupling effect.

\section{Modified DGP-model}
To solve this problem let us add to the action (\ref{DGPaction})
the following terms
\begin{equation}\label{addAction}
S_{add}=M_{*}^{3}\int_{y=0}\varphi(x) \tilde R\sqrt{-\tilde
g}\,d^{4}x + \int \Phi(x,y) R^{2}\sqrt{-g}\,d^{4}xdy.
\end{equation}
The first term in (\ref{addAction}) corresponds to the
four-dimensional Brans-Dicke theory with $\omega=0$ (where
$\omega$ is the Brans-Dicke parameter). The absence of the kinetic
terms for the fields $\varphi(x)$ and (especially) $\Phi(x,y)$
look rather strange. Nevertheless one can recall that SUSY is
based on the use of such "auxiliary"\, fields, which are necessary
for implementing the supersymmetry transformations. A simple
example with the fields of such type in classical field theory can
be found in \cite{Ramon}. In any case, there are no strong
objections against considering fields of this type, although these
fields will be used for other purposes, not that as in SUSY. We
also assume that $\varphi(x)\equiv 0$ in the background, otherwise
the redefinition of $\Omega$ in equation (\ref{DGPaction}) is
needed.

From the equation of motion for the field $\Phi(x,y)$ we have the
following condition for the five-dimensional curvature
\begin{equation}\label{5curv}
R=0.
\end{equation}
This equation holds in the bulk.

Equations (\ref{mu-nu}), (\ref{mu-4}), (\ref{4-4}) and
(\ref{contracted-44}) are modified by the terms of
(\ref{addAction}) and take the form (it is easy to see that one
can use the same gauge condition for the field $h_{\mu 4}$ as that
used in the previous section):

1) $\mu\nu$-component
\begin{eqnarray}\label{mu-nu2}
 & &\frac{1}{2} \left(\Box h_{\mu\nu}-
\partial_\mu \partial^\rho
h_{\rho\nu}-\partial_\nu \partial^\rho h_{\rho\mu} +
\partial_4\partial_4 h_{\mu\nu}+\partial_\mu
\partial_\nu h+ \partial_\mu \partial_\nu
h_{44}\right) +\\ \nonumber &+& \frac{1}{2}
\gamma_{\mu\nu}\left(\partial^\rho
\partial^\sigma h_{\rho\sigma}-\Box
h - \partial_4\partial_4  h - \Box h_{44} \right) +\\ \nonumber
&+& \frac{\Omega}{2}\delta (y)\left[\left(\Box h_{\mu\nu}-
\partial_\mu \partial^\rho
h_{\rho\nu}-\partial_\nu \partial^\rho h_{\rho\mu}+\partial_\mu
\partial_\nu h\right)+\gamma_{\mu\nu}\left(\partial^\rho
\partial^\sigma h_{\rho\sigma}-\Box
h\right)\right]+\\ \nonumber &+&
\frac{1}{\hat\kappa}\left(\partial_{\mu}\partial_{\nu}\varphi -
\eta_{\mu\nu}\Box\varphi\right)\delta(y)= -\frac{\hat \kappa}{2}
t_{\mu\nu}(x)\delta(y),
\end{eqnarray}
(corresponding equations for the Brans-Dicke theory can be found,
for example, in \cite{Thorn}),

 2) $\mu 4$-component
\begin{equation}\label{mu-42}
\partial_4 ( \partial_\mu h - \partial^\nu  h_{\mu\nu})= 0,
\end{equation}

 3) $4 4$-component
\begin{equation}\label{4-42}
\partial^\mu \partial^\nu  h_{\mu\nu} - \Box  h =0,
\end{equation}

4) auxiliary equation
\begin{equation}\label{contracted-442}
\partial_4\partial_4 h + \Box h_{44} + \frac{2}{\hat\kappa}\,\Box\varphi\delta(y) =
\frac{\hat \kappa}{3}\, t(x)\delta(y),
\end{equation}
The second term in (\ref{addAction}) does not contribute to the
equations since it contains $R$ squared. Thus the main purpose of
this term is to make the five-dimensional curvature equal to zero
in the bulk. It is evident that equation of motion for the field
$\varphi(x)$ does not contradict (\ref{4-42}). Thus only the case
$\omega=0$ is compatible with action (\ref{DGPaction}), otherwise
the existence of kinetic term for $\varphi(x)$ leads to the
condition $\Box\varphi=0$ and this field decouples from the
theory. Linearizing equation (\ref{5curv}) and taking into account
(\ref{4-42}), we get
\begin{eqnarray}\label{5curvLin}
\partial_4\partial_4  h + \Box h_{44}=0.
\end{eqnarray}
It means that
\begin{equation}\label{var}
\Box\varphi = \frac{\hat \kappa^2}{6}\, t.
\end{equation}
Integrating (\ref{5curvLin}) in the gauge
$h_{44}(x,y)=e^{-k|y|}\phi(x)$ in the limits $(-\infty,\infty)$
and using the physical boundary conditions for the field
$h_{\mu\nu}$, we get
\begin{equation}
\Box\phi=0.
\end{equation}
We see, that $h_{44}$-field does not interact with matter on the
brane, i.e. it decouples from the effective theory. This also
means that the linear approximation is valid. Moreover, we can
totally gauge out this field with the help of $\xi_{4}$ of the
form
\begin{equation}\label{xi4_1}
\xi_4(x,y)=\frac{1}{4}\int_{-y}^y h_{44} (x,y') dy'
\end{equation}
instead of (\ref{xi4}). Analogously to what was made in the
previous section, we have the following conditions for the field
$h_{\mu\nu}$:
\begin{equation}\label{T-1}
h =0,
\end{equation}
\begin{equation}\label{gaugecond-1}
 \partial^\mu  h_{\mu\nu} = 0.
\end{equation}

Let us explain what has happened. The field $\varphi$ takes over
the role of the $h_{44}$-component of metric fluctuations. Since
this field does not exist "inside"\, the curvature, one does not
need to worry about its absolute value -- it does not affect the
validity of linear approximation. At the same time the field
$\Phi$ makes $h_{44}$ not interacting with matter on the brane
(one can check that in the absence of the corresponding term it is
not necessarily so).

Now we are ready to find the equation for the $\mu\nu$-component.
In gauge (\ref{T-1}), (\ref{gaugecond-1}) and subject to
(\ref{var}) it takes the well-known form
\begin{eqnarray}\label{mu-nu3}
\Box h_{\mu\nu}+\partial_4\partial_4 h_{\mu\nu} + \Omega\delta
(y)\Box h_{\mu\nu}= -\hat \kappa
\delta(y)\left(t_{\mu\nu}-\frac{1}{3}\left[\eta_{\mu\nu}-\frac{\partial_{\mu}\partial_{\nu}}
{\Box}\right]t\right).
\end{eqnarray}
It coincides with that derived in \cite{DGP} and was examined in a
lot of papers, so we will not discuss it here. All the predictions
obtained with the help of this equation are valid in this case
too.

\section{Discussion and final remarks}
In this paper we discussed the DGP-model with additional terms,
which include two "auxiliary"\ scalar fields. The first one takes
over the role of the $h_{44}$-component of metric fluctuations
(the radion field), which is indispensable for the original
DGP-model and makes the equations consistent, whereas the second
one is responsible for dropping out this component from the
theory. Thus, the strongly interacting scalar mode vanishes away
as well as the strong coupling effect. The effective equation for
the $\mu\nu$-component of metric fluctuations appears to be the
same, as in the case of the original DGP setup. Also it is not
difficult to check that non-linear term $\sim\varphi\tilde
R_{\mu\nu}$, which appears in the Einstein equations, can be
neglected in comparison with the term $\sim\Omega\tilde
R_{\mu\nu}$ even for such massive objects like those in our Solar
system.

Now let us discuss the problem of ghosts in the model. First, we
note, that the four-dimensional Brans-Dicke theory with $\omega=0$
is stable, which can be shown with the help of conformal
rescaling. It is reasonable to suppose, that this property
survives in the five-dimensional theory too. Second, the field
$\varphi(x)$ behaves exactly as the radion at the linear order
(compare (\ref{mu-4-1}), (\ref{4-4-1}), (\ref{contracted-44-1})
and (\ref{mu-42}), (\ref{4-42}), (\ref{contracted-442})), which is
not a ghost in the original DGP-model. And third, the existence of
ghosts lead to additional repulsive forces, which are absent in
the case under consideration. This formal argumentation can be
used in favor of the absence of ghosts at least in the second
variation Lagrangian.

In this paper we solve equations of motion only at the linear
order. One can ask about the possibility of the presence of the
strong coupling effect at the next orders, for example, through
the cubic and quartic terms of the effective Lagrangian. To this
end it is necessary to make a thorough analysis analogous to the
one, which was made in \cite{Rub}. Nevertheless the strong
coupling effect manifests itself even at the linear order, as it
was shown in Section~2. This can be also seen through the
diagonalization of the second variation Lagrangian of the model.
In the original DGP-model such procedure leads to appearance of
the kinetic term for the radion with a small normalization
constant, leading to the strong coupling at the next orders (see
\cite{Luty,Rub}). As it was shown above, the field $\varphi(x)$
acts in the same way as the radion (at the linear order).
Nevertheless, these fields have different nature, which becomes
apparent after diagonalization of the Lagrangian. In the
four-dimensional theory diagonalization of the Brans-Dicke action
by conformal rescaling results in the appearance of the scalar
field, interacting directly with matter with the same coupling
constant as the tensor gravity. The linearized substitution for
the $\mu\nu$-component of metric fluctuations, which corresponds
to the conformal rescaling, does not contain large pieces, which
can lead to the strong coupling at the next orders. In the case of
the model described in Section~3 the corresponding diagonalization
is not so trivial, as that in the four-dimensional case.
Nevertheless naive calculations show that situation appears to be
analogous to the four-dimensional case: the main contributions to
the kinetic term of the field $\varphi(x)$ come from the brane
action, and these contributions appear to be much larger than
those coming from the five-dimensional curvature. At the same time
the field $\varphi(x)$ can appear, for example, in the cubic terms
of the action only through the substitution for the
$\mu\nu$-component, which seems not to contain large pieces,
contrary to the case of original DGP-setup, in which analogous
substitution does contain large pieces (see \cite{Rub}). Moreover,
the radion also "lives"\ in the five-dimensional curvature and can
appear in the cubic order terms "independently"\ of the
substitution. In other words, the four-dimensional Brans-Dicke
gravity with $\omega=0$ dominates on the brane at the small
distances, instead of the four-dimensional gravity with
contribution of the radion in the original DGP-model. If there
were no radion, which leads to the strong coupling, in the
original DGP-setup, the only pure four-dimensional gravity would
live on the brane at the small distances. After the modification
of the model the radion drops out from the theory, whereas the
four-dimensional Brans-Dicke gravity takes over the role of the
"pure four-dimensional gravity" of the original DGP-model. Thus,
there seem to be no prerequisites for the strong coupling effect,
coming from the second variation Lagrangian, contrary to the case
of the original DGP-model.

However, since the absence of the strong coupling effect was shown
only at the level of solving the classical linear equations of
motion, a further thorough analysis is necessary. This analysis
can also clarify the fate of the radion and the $\mu4$-component
of metric fluctuations in the model.

One can worry about the fact that the field $\Phi$ is not defined
by the equations of motion. This problem can be easily solved by
adding an extra term to the action, for example, one can choose
\begin{equation}
S_{\Phi}=\frac{\alpha}{3}\int \Phi^{3}(x,y)\sqrt{-g}\,d^{4}xdy,
\end{equation}
where $\alpha$ is real and positive. In this case equation
(\ref{5curv}) will take the following form:
\begin{equation}
R^{2}+\alpha\Phi^{2}=0,
\end{equation}
which uniquely determines the field $\Phi$ ($\Phi\equiv 0$).

Here we do not consider the problem of the incorrect tensor
structure of the graviton, which is analogous to the
vDVZ-discontinuity in massive gravity \cite{vDV,Zakh}. This issue
was discussed in a lot of papers including \cite{DGP}. It seems
that this problem is inherent at least to five-dimensional models
with flat background and matter localized on the brane (see
equation (\ref{4-42})). It is reasonable to suppose that it can be
solved in models with warped geometry. Hence the modification of
the DGP-model proposed above does not make the model applicable
for describing gravity in our world. Nevertheless one can hope
that the use of additional fields non-minimally coupled to gravity
(not necessarily without kinetic terms) can be quite efficient in
some more realistic models, which admit long-range modification of
gravity and suffer from the problem of strong coupling effect.

\bigskip
{\large \bf Acknowledgments}
\medskip \\
The author is grateful to M.V.~Libanov and I.P.~Volobuev for
valuable discussions. The work was supported by the RFBR grant
04-02-16476, by the grant UR.02.02.503 of the scientific program
"Universities of Russia" and by the grant NS.1685.2003.2 of the
Russian Federal Agency for Science.

\end{document}